\documentclass[oldversion]{aa}
\usepackage{txfonts,epsfig,graphicx}

\def\tab#1{{Table\ }\ref{tab:#1}}
\def\fig#1{{Fig.\ }\ref{fig:#1}}
\newcounter{Tableau}
\newcounter{Figure}
\renewcommand{\thetable}{\arabic{Tableau}}
\renewcommand{\thefigure}{\arabic{Figure}}

\begin{document}

\title{Probing isotopic ratios at z = 0.89: \\ molecular line absorption in front of the quasar PKS~1830-211}

\author{Muller S. \inst{1,2} \and Gu\'elin M. \inst{2} \and Dumke M. \inst{3,2} 
\and Lucas R. \inst{2} \and Combes F. \inst{4}}
\offprints{muller@asiaa.sinica.edu.tw}
\institute{
Academia Sinica Institute of Astronomy and Astrophysics (ASIAA), P.O. Box 23-141, Taipei, 106 Taiwan
\and
Institut de RadioAstronomie Millim\'etrique (IRAM), 300 rue de la piscine, F-38406 St Martin d'H\`eres, France
\and 
European Southern Observatory (ESO), Alonso de Cordova 3107, Casilla 19001, Santiago 19, Chile
\and
Observatoire de Paris, LERMA, 61 av. de l'Observatoire, F-75014 Paris, France}

\date {Received 13 March 2006 / Accepted 26 July 2006}
%\thesaurus{}

\titlerunning{Probing isotopic ratios at z = 0.89}
\authorrunning{Muller et al. 2006}

\abstract{With the Plateau de Bure interferometer, we have measured
the C, N, O and S isotopic abundance ratios in the arm of a spiral
galaxy with a redshift of 0.89. The galaxy is seen face-on according
to HST images. Its bulge intercepts the line of sight to the
radio-loud quasar \object{PKS~1830-211}, giving rise at mm wavelengths to two
Einstein images located each behind a spiral arm. The arms appear in
absorption in the lines of several molecules, giving the opportunity
to study the chemical composition of a galaxy only a few Gyr old. The
isotopic ratios in this spiral galaxy differ markedly from those
observed in the Milky Way. The $^{17}$O/$^{18}$O and $^{14}$N/$^{15}$N
ratios are low, as one would expect from an object too young to let
low mass stars play a major role in the regeneration of the gas.

}
\maketitle

\section{Introduction}

The C, N, O, S isotopic abundances are enlightening relics of the interstellar
medium (ISM) past history. Not only are these elements the most abundant ones
after H and He, but they are those whose synthesis in stars has been the best
investigated. The CNO isotopes are tentatively classified into primary elements
($^{12}$C, $^{16}$O,...), formed directly by He burning, and secondary ones
($^{13}$C, $^{14}$N, $^{17}$O,...) formed at a later stage from primary elements,
e.g. through the CNO cycles. In simple evolution models, the fraction of
secondary elements increases with time. Actually, chemical evolution is not so
straightforward and we know that isotopes like $^{14}$N and perhaps $^{13}$C are
already in part produced in the first generation of stars.
 
Whereas isotopic abundances can be precisely measured in the Sun and on the
Earth, their values in the Galactic ISM and, mostly, in the extragalactic
ISM are hard to derive. In a pioneering work, Penzias and co-workers
(Penzias et al. \cite{pen80}, \cite{pen81}) used the millimeter (mm) emission lines of molecular
isotopologues to infer the isotopic ratios in a dozen molecular sources.
They found significant differences between the Solar System, the local ISM,
the Galactic Center and circumstellar envelopes that they interpreted in terms
of degree of nuclear processing, the Solar System representing the state of
the local ISM some 4.5 Gyr ago. This picture of the Galactic ISM remains
essentially valid today, although it has been recognized that the molecular
ratios involving D and $^{13}$C are affected by isotopic fractionation or
selective photodissociation and may not reflect the elemental isotopic
abundance ratios (Wilson \& Matteucci \cite{wil92}).

So far, only a few isotopic abundance ratios have been reported in external
galaxies. Those are mostly the C, N, O and S isotopic ratios derived in the
nuclei of four nearby starburst galaxies, M~82, IC~342, NGC~253 and NGC~4945,
from the mm emission of CO, CS and HCN (Henkel \& Mauersberger \cite{hen93},
Henkel et al. \cite{hen98}, Chin et al. \cite{chi99b}, Wang et al. \cite{wan04}, Mart\'in et al. \cite{mar05}).
These ratios are very uncertain due to beam dilution, because the lines of
the rare isotopologues are weak, and because the line opacity of the main isotopologues
is unknown. Moreover, the studies deal with nuclear regions that may
not reflect the disk ISM abundances, as is the case in the Milky Way.

Another powerful way of studying the composition of the ISM consists in
observing the molecular lines in absorption. The method needs only a strong
background source and is insensitive to distance and to primary beam dilution.
It yields directly the line opacities, provided the background source is fully
covered by the intercepting clouds (e.g. Lucas \& Liszt \cite{luc98}). As far as
molecular isotopic ratios are concerned, the method is limited to the line of
sight to the brightest radio galaxies and quasars intercepted by deep absorption
line systems. Few such absorption systems have been identified so far at
millimeter wavelengths. They arise in the Milky Way (Lucas \& Liszt \cite{luc96}, \cite{luc98})
and in a few distant galaxies with redshifts in the range z = 0.2 -- 0.9
(Wiklind \& Combes \cite{wik94}, \cite{wik95}, \cite{wik96}, Combes \& Wiklind \cite{com98}, Combes \cite{com99}).

In this article, we present high sensitivity absorption spectra arising from 
the most remote presently known system, located at z = 0.89. The spectra cover the lines of a dozen
of molecular species, including the rare isotopomers HC$^{17}$O$^+$, H$^{15}$NC
and C$^{34}$S. From these spectra, we derive the C, N, O and S isotopic ratios
in the spiral arm of a young, seemingly normal galaxy.\footnote{We adopt the
usual convention to simply denote by C, N, O and S the abundant isotopes
$^{12}$C, $^{14}$N, $^{16}$O and $^{32}$S.}

\section{The gravitational lens towards PKS~1830-211}

PKS~1830-211 is a radio loud quasar with a redshift of z = 2.5 (Lidman et al. \cite{lid99}),
whose line of sight is intercepted by at least two galaxies: a relatively nearby
one, responsible for narrow HI line absorption at redshift z = 0.19 (Lovell et al.
\cite{lov96}), and a more distant one that gives rise to broad molecular and HI line absorption
at z = 0.89 (Wiklind \& Combes 1996 and 1998, hereafter \cite{wc96} \& \cite{wc98},
G\'erin et al. \cite{ger97}, Menten et al. \cite{men99}, Chengalur et al. \cite{che99},
Muller \& Gu\'elin \cite{mul03}). No molecular absorption is detected in the first galaxy
(\cite{wc98}) while the latter galaxy acts as
a gravitational lens and gives rise, at radio wavelengths, to two compact sources
(NE and SW) embedded in a faint Einstein ring (see Fig.3 from Chengalur et al. \cite{che99},
see also Jauncey et al. \cite{jau91}).
The ring has a steep spectral index and fades away at short wavelengths
(Carilli et al. \cite{car98}), so that at mm wavelengths, the quasar image reduces essentially
to the NE and SW sources that are distant by $\simeq$ 1\arcsec, plus two weak 'tails'
(Frye et al. \cite{fry97}, \cite{wc98}, Carilli et al. \cite{car98}). 

The z = 0.89 absorption system shows two main velocity components: a
broad one at z = 0.88582 (assuming V = 0 kms$^{-1}$ in heliocentric
coordinates), associated with the SW source, and a narrow component,
147 kms$^{-1}$ lower in velocity, associated with the NE source
(Frye et al. \cite{fry97}, \cite{wc98}). Although both components are detected in
absorption in HI and in molecular lines, the NE 
component is more conspicuous in the HI line and the SW one in the
molecular lines (\cite{wc98}, Chengalur et al. \cite{che99}).

Recently, Winn et al. (\cite{win02}) have published a HST I-band image of the z = 0.89 lensing
galaxy. This latter appears as a typical nearly face-on spiral (probably of type Sb or Sc).
The low inclination of the galaxy is confirmed by Koopmans \& de Bruyn (\cite{koo05}), who modelled
its kinematics from the HI absorption profile and find i = 17\degr -- 32\degr. 
The bulge of the galaxy is located near the
center of the radio ring. The SW radio source appears to lie right on
top of a spiral arm, at a (model dependent) distance of $\simeq$ 2 kpc
from the center. The NE source lies at about twice this distance
($\simeq$ 4 kpc) on the other side of the nucleus. Both absorption
components seem thus to arise in the counterpart of what we call in the
Milky Way the Molecular Ring.

Adopting the cosmological parameters H$_0$ = 70 kms$^{-1}$
Mpc$^{-1}$, $\Omega_M$ = 0.3 and $\Omega_\Lambda$ = 0.7, the redshift
z = 0.89 corresponds to a lookback time of 7.2 Gyr, yielding an age
$\leq$ 6 Gyr for the galaxy. Assuming that the first generation of stars
formed 1 Gyr after the Big Bang, stars with masses $<$ 1.5 M$_\odot$ barely
had the time to contribute to the enrichment of the interstellar medium.

\begin{table}[h] \setcounter{Tableau}{1}
\caption{Basic data for PKS~1830-211 and the intervening galaxy.} \label{tab:pks}
\begin{center} \begin{tabular}{llll}
\hline \hline
\multicolumn{4}{l}{PKS~1830-211} \\
\hline
NE image          & R.A. (J2000) & 18$^\mathrm{h}$33$^\mathrm{m}$39\fs 932  & (1) \\
                  & Dec. (J2000) & $-$21\degr 03\arcmin 39 \farcs 73       & (1) \\

SW image          & R.A. (J2000) & 18$^\mathrm{h}$33$^\mathrm{m}$39\fs 886  & (1) \\
                  & Dec. (J2000) & $-$21\degr 03\arcmin 40\farcs 45       & (1) \\

Redshift          &             & z = 2.5                 & (2) \\

3mm flux density  & (variable)  & 1.5 -- 2.5 Jy            &     \\
Flux distribution & NE image    & $\sim$ 63\%             & (3) \\
                  & SW image    & $\sim$ 37\%             & (3) \\
\hline
\multicolumn{4}{l}{Intervening galaxy} \\
\hline
Redshift          &               & z = 0.88582           & (3) \\
Type              &               & Sb or Sc              & (4) \\
Inclination       &               & 17\degr -- 32\degr     & (5) \\
Linear scale      &               & 1\arcsec \  $\sim$ 7.8 kpc    & \\
Velocity          & SW absorption & $\sim$ 0 kms$^{-1}$         & (6) \\
                  & NE absorption & $\sim$ $-$147 kms$^{-1}$      & (6) \\
Age               &               & $\leq$ 6 Gyr          & \\
\hline
\end{tabular} \end{center}
\mbox{\,} %\vskip -.8cm
References: (1) Subrahmanyan et al. \cite{sub90}; (2) Lidman et al. \cite{lid99};
(3) \cite{wc96}; (4) Winn et al. \cite{win02}; (5) Koopmans \& de Bruyn (\cite{koo05}); (6) \cite{wc98}.
\end{table}

\section{Observations} \label{obs}  

With the Plateau de Bure Interferometer (PdBI), we have carried out a
high sensitivity and high spectral resolution survey of a dozen of mm
absorption lines in the line of sight to PKS~1830-211. This survey is
the result of a number of 2 -- 4 hour-long observing sessions, carried out
between 1999 and 2005 (see Table 2). The antennas were most of the time in a compact
configuration which led to a synthesized beam $\sim$10\arcsec \ at $\nu$ =
100 GHz ($\delta$ = $-$21\degr), much larger than the source size.
The phase reference was set at R.A. = 18$^\mathrm{h}$33$^\mathrm{m}$39\fs 937, Dec. =
$-$21\degr03\arcmin39\farcs80 (J2000). The spectral resolution was 1.25 MHz before
2000, and 2 or 4 times better afterwards. For calibration
purpose, a strong continuum source (usually 3C273 or 3C345), free of
any absorption at the frequencies of interest was observed for 10
minutes shortly before and after PKS~1830-211.
 
The data were reduced and analyzed with the GILDAS/CLIC software. The
data from each observing session were first calibrated for
instrumental RF bandpass by fitting a high order polynomial on
the spectrum observed for the strong continuum source. We checked
that this calibration was accurate to better than 1\%.

\onltab{2}{
\begin{table*}[t] \setcounter{Tableau}{2}
\caption{Observational data.} \label{tab:obsdata}
\begin{center} \begin{tabular}{lrllll}
\hline \hline 
                              & \multicolumn{1}{c}{Redshifted} &                            & \multicolumn{1}{c}{Time} &        \\ 
\multicolumn{1}{c}{Line} & \multicolumn{1}{c}{frequency}  & \multicolumn{1}{c}{Date} & \multicolumn{1}{c}{ON}     & \multicolumn{1}{c}{Array$^\diamond$}\\
                              & \multicolumn{1}{c}{$^\dagger$ (MHz)}      &                            & \multicolumn{1}{c}{source}      &           \\

\hline
HCO$^+$ ($2\leftarrow 1$) & 94587.58 & 1999/06/11 & 3.0 h & 4C1  \\          
              &          & 2001/03/20 & 2.5 h & 4ANT  \\
              &          & 2001/06/29 & 3.5 h & 4ANT \\
              &          & 2002/06/25 & 3.0 h & 5D  \\
              &          & 2003/02/26 & 2.2 h & 6Ap  \\
              &          & 2003/03/01 & 2.6 h & 6Ap   \\
\hline
H$^{13}$CO$^+$ ($2\leftarrow 1$) & 92006.01 & 1999/06/13 & 2.9 h & 4C1  \\   
                     &          & 2002/06/21 & 2.7 h & 5D   \\
                     &          & 2002/06/23 & 3.7 h & 5D  \\
                     &          & 2002/07/15 & 1.3 h & 5D   \\
\hline
HC$^{18}$O$^+$ ($2\leftarrow 1$) & 90317.61 & 1999/08/03 & 3.2 h & 4ANT  \\
                     &          & 1999/09/09 & 2.8 h & 4C2  \\   
                     &          & 1999/10/02 & 2.8 h & 4C2  \\
                     &          & 1999/11/12 & 1.8 h & 5D  \\
                     &          & 2001/05/08 & 1.7 h & 4ANT  \\
\hline
HC$^{17}$O$^+$ ($2\leftarrow 1$) & 92327.67 & 1999/07/08 & 4.3 h & 4D2  \\   
                     &          & 1999/07/15 & 3.7 h & 4D2  \\
                     &          & 1999/07/25 & 3.0 h & 4D2  \\
                     &          & 1999/09/24 & 2.0 h & 4C2  \\
                     &          & 1999/10/14 & 2.3 h & 4B1  \\
                     &          & 1999/10/26 & 2.0 h & 4B1  \\
                     &          & 1999/10/29 & 3.2 h & 4D1  \\
                     &          & 1999/11/01 & 3.2 h & 4D1  \\
                     &          & 1999/11/08 & 1.7 h & 5D   \\
\hline
HCN ($2\leftarrow 1$) & 93996.83 & 1999/08/04 & 2.0 h & 4ANT  \\             
          &          & 2003/08/03 & 1.3 h &  30M  \\
\hline
H$^{13}$CN ($2\leftarrow 1$) & 91566.38 & 1999/08/29 & 2.7 h & 5D  \\        
                 &          & 2001/07/03 & 2.6 h &  5D  \\
\hline
HC$^{15}$N ($2\leftarrow 1$) & 91264.34 & 1999/08/25 & 4.2 h & 4D1  \\       
                 &          & 1999/10/04 & 1.6 h & 4C2  \\
                 &          & 2001/05/28 & 2.0 h & 4ANT  \\
                 &          & 2001/05/30 & 1.5 h & 4ANT  \\
                 &          & 2001/08/13 & 3.3 h & 4D1  \\
                 &          & 2001/08/14 & 2.8 h & 4D1  \\
                 &          & 2001/08/15 & 3.7 h & 4D1  \\
\hline
HNC ($2\leftarrow 1$) & 96151.68 & 2002/08/07 & 3.7 h & 5D  \\               
\hline
HN$^{13}$C ($2\leftarrow 1$) & 92362.69 & 2001/08/26 & 2.7 h & 4D1  \\       
                 &          & 2001/10/15 & 1.2 h & 4ANT  \\
\hline
H$^{15}$NC ($2\leftarrow 1$) & 94244.99 & 2001/08/04 & 2.8 h & 4D1  \\       
                 &          & 2001/08/07 & 3.3 h & 4D1  \\
                 &          & 2001/08/25 & 2.7 h & 4D1  \\
\hline
DNC ($2\leftarrow 1$) & 80924.89 & 2001/08/28 & 2.7 h & 4D1  \\              
\hline
CS ($4\leftarrow 3$) & 103909.33 & 1999/07/09 & 3.3 h & 4D2  \\              
\hline
C$^{34}$S ($4\leftarrow 3$) & 102246.49 & 1999/09/21 & 1.8 h & 4C2  \\       
                &           & 1999/09/23 & 2.2 h & 4C2  \\
                &           & 2001/05/09 & 1.8 h & 3ANT  \\
                &           & 2001/05/10 & 3.4 h & 5D  \\
                &           & 2001/05/14 & 2.6 h & 5D  \\
\hline
H$_2$S ($1_{10} \leftarrow 1_{01}$)$^\ddagger$ & 89490.39 & 2005/07/05 & 4.3 h & 5D  \\      
                           &          & 2005/07/29 & 3.2 h & 5D  \\
                           &          & 2005/09/05 & 3.0 h & 5D  \\
                           &          & 2005/09/09 & 1.8 h & 5D  \\
\hline
\end{tabular} \end{center}
\mbox{\,} %\vskip -.8cm
Notes: \\
$^\diamond$ The number of antennae is given with the name of the array configuration if defined. \\
$^\dagger$ Redshifted frequencies are given for z = 0.88582. \\
$^\ddagger$ The H$_2^{34}$S (1$_{10} \leftarrow 1_{01}$) (89038.47 MHz) and H$_2^{33}$S (1$_{10}$-1$_{01}$) (89256.88 MHz) were observed simultaneously in the same correlator setup.
\end{table*}  
}

Next, the spectral data were calibrated in amplitude and phase
with respect to the PKS~1830-211 continuum, whose flux was $\simeq$ 2 Jy
at $\lambda$ = 3 mm. PKS~1830-211 being unresolved in the
compact antenna configurations, the phases in this process were
referenced to the barycenter of the continuum emission and the
amplitudes normalized to the total continuum flux. Finally, the
spectra observed on the different sessions and with the different 
baselines were co-added and a third-degree baseline withdrawn from 
the global spectrum, in order to remove small RF bandpass residuals (amplitude $\sim$ 2\%) caused by
instrumental drifts and/or strong atmospheric fluctuations.

The total integration time per line ranges from 1 h to 20 h; the rms noise on
the weakest lines is between 2 and 7 mJy per 1.25 MHz channel.

In addition to the compact configuration observations just described,
and which served as the basis of our molecular line survey, we
observed on 2003 February 26$^{th}$ and March 1$^{st}$ the HCO$^+$
($2\leftarrow 1$) line in the extended A configuration. This configuration
includes 400 m-long E-W and NE-W baselines that allow to spatially
resolve the NW and SW source components. We used a special procedure
to reduce these data. This procedure is described in \S\ref{position}.

Finally, we also observed on 2003 August 3$^{rd}$ the ($2\leftarrow 1$) line of HCN with
the IRAM 30-m telescope. The spectrum shown in \fig{spectra-strong} is the sum
of the 30-m and PdBI data.

\section{Data analysis}

\subsection{Position of the absorptions} \label{position}

Using the BIMA array, Frye et al. \cite{fry97} (see also Swift et al. \cite{swi01})
successfully resolved the two continuum components. They derived
an angular separation of 0.99\arcsec \ $\pm$ 0.05\arcsec. Further evidence that
the NE component was indeed responsible for the second absorption at
V = $-$147 kms$^{-1}$ was shown by \cite{wc98}. They used the BC configuration
of the PdBI and fitted the position of the barycenter of emission
directly from the visibilities.

Our A configuration PdBI observations of the HCO$^+$ ($2\leftarrow 1$) line,
which combine high sensitivity and high angular resolution enable
us to locate with a much higher relative accuracy the velocity
components. For this, we used the following procedure. 

The RF bandpass was first calibrated using the quasar 1749+096 (on
2003 Feb. 26$^{th}$) or 3C273 (on 2003 Mar. 1$^{st}$). Next, the
continuum data on PKS~1830-211 were self-calibrated by calculating the
complex gains (amplitude and phase) corresponding to a point-like
source of flux unity at the phase center. The latter was fixed through
this procedure at the barycenter of the continuum emission. Then, the
calculated gains were applied to the line visibilities and the continuum
was subtracted. The resulting visibilities were fitted by a point source,
channel by channel, to trace the position of the absorption. At this step,
the absorptions at V$\simeq -$147 kms$^{-1}$ and V$\simeq$0 kms$^{-1}$ were already
clearly found to arise from two different locations (NE and SW respectively)
separated by 1.01\arcsec \ $\pm$ 0.03\arcsec \ at a position angle of $\sim$42\degr.

\begin{figure}[h] \addtocounter{Figure}{1} \resizebox{\hsize}{!}
{\includegraphics {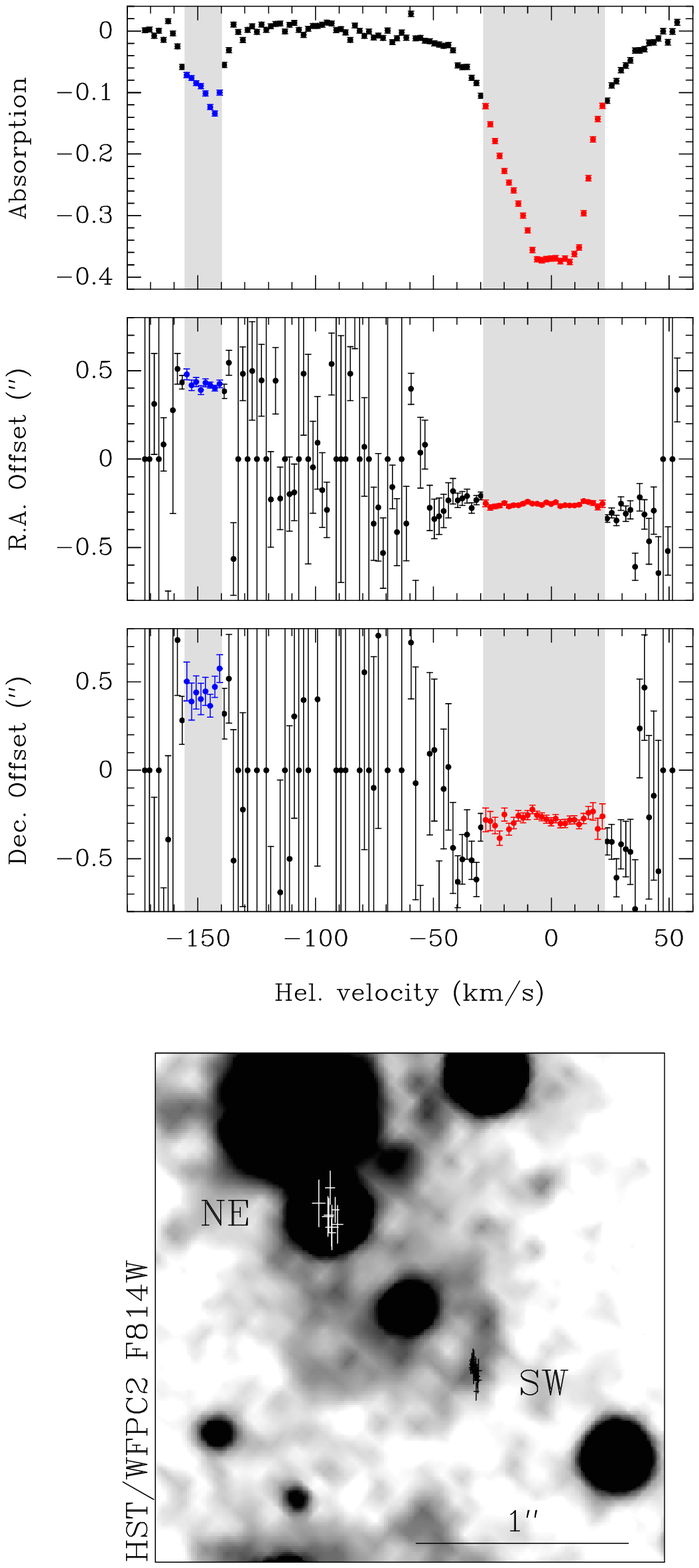}}
\caption{Positions of the SW and NE absorption components as observed with
PdBI in the HCO$^+$ ($2\leftarrow 1$) line. The R.A. and Dec. offsets (top)
at each velocity channels, indicating the position of the absorption (shaded area),
were obtained from a fit of the interferometric visibilities. These positions are
reported on a Hubble Space Telescope I band image of the z = 0.89 galaxy (bottom).
The HST data were retrieved from the STScI Archive site:
''http://archive.stsci.edu/hst/search.php'' and reduced within IRAF.
The SW absorption clearly falls right on top of a spiral arm.}
\label{fig:uvfit}
\end{figure}

The whole process was iterated: 
{\it i}) A new input continuum model, consisting of two point-like sources,
located at the previously derived NE and SW positions and with fluxes in the
ratio 1.7 (see \tab{magfactor}), was used to self-calibrate the data.
{\it ii)} the new continuum was subtracted from the calibrated visibilities
to yield the amplitude of the absorption and its position in each velocity
channel (\fig{uvfit}). 

We confirm that the two main absorption components (around V = 0 kms$^{-1}$
and V = $-$147 kms$^{-1}$) arise from two distinct compact sources, whose
relative positions, estimated from weighted averages over each component, 
are (0.420\arcsec \ $\pm$ 0.008\arcsec; 0.45\arcsec \ $\pm$ 0.03\arcsec) for the NE component and
($-$0.255\arcsec \ $\pm$ 0.002\arcsec; $-$0.279\arcsec \ $\pm$ 0.006\arcsec) for the SW component.
This corresponds to a separation of 0.99\arcsec \ $\pm$ 0.03\arcsec \ and a position angle
of 43\degr $\pm$1\degr, in good agreement with the separation and position
angle of the continuum sources derived from BIMA, MERLIN and VLA observations.

One purpose of our extended configuration observations was to check
whether the 80 kms$^{-1}$-broad absorption component centred at 0
kms$^{-1}$ was arising entirely from the compact SW image, or if
the line shoulders and/or wings were arising from the dimmer `tails' and 
`knots' visible at longer wavelengths, or from the weak Einstein ring. 
It should be noted that the V = 0 kms$^{-1}$ HCO$^+$ and HCN absorption
profiles observed at 3-mm at low angular resolution are almost twice broader 
than those observed at 6-mm with a 0.1\arcsec \ resolution (Carilli et al. \cite{car98}).  
 
\fig{uvfit} shows that the absorption arises within 0.1\arcsec \ (0.2\arcsec \ in
weak wings of the line) from a point-like source and shows no obvious
velocity/position gradient. In particular, the emission at V $<$ $-$20 kms$^{-1}$ 
and V $>$ 20 kms$^{-1}$, which is not observed in the high resolution VLA and 
VLBA observations, does not arise from the `tails' and `knots', but comes 
from the compact SW `core'.   

\subsection{Magnification ratio}

The flux of PKS~1830-211 is known to be variable. For example, van Ommen et al. 
(\cite{van95}) report a variation of the 15 GHz flux by a factor of 2 between 1990
and 1991, whereas Wiklind \& Combes (\cite{wik99}) observe a similar increase of the
3-mm flux between 1997 and 1998. Although the 3-mm flux was mostly stable and
close to 2.5 Jy in the course of our survey, we noted 20 -- 30\% changes between 
1999 and 2001 and between 2001 and 2002. There was no significant change around 
August 2002, the period when the HNC line was observed.

A question central to this study is whether the magnification ratio of the NE
and SW sources, $\Re$ = NE/SW, is also variable, as this would affect the amplitude
of the V = 0 and $-$147 kms$^{-1}$ absorption components with respect to the
continuum flux. $\Re$ is expected to vary after a sudden increase of the flux
of the background quasar, since the increase will first appear in the NE source,
then in the SW source after a delay $\Delta \tau$. The path length from the quasar
to the observer is indeed shorter for the rays passing through the NE source
(Nair et al. \cite{nai93}). $\Delta \tau$ has been measured by Lovell et al. (\cite{lov98}) and by
Wiklind \& Combes (\cite{wik99}) and found to be about 25 days. Micro-lensing
events caused e.g. by stars may also randomly affect $\Re$ for short periods of time.

\begin{figure}[h] \addtocounter{Figure}{1} \resizebox{\hsize}{!}
{\includegraphics{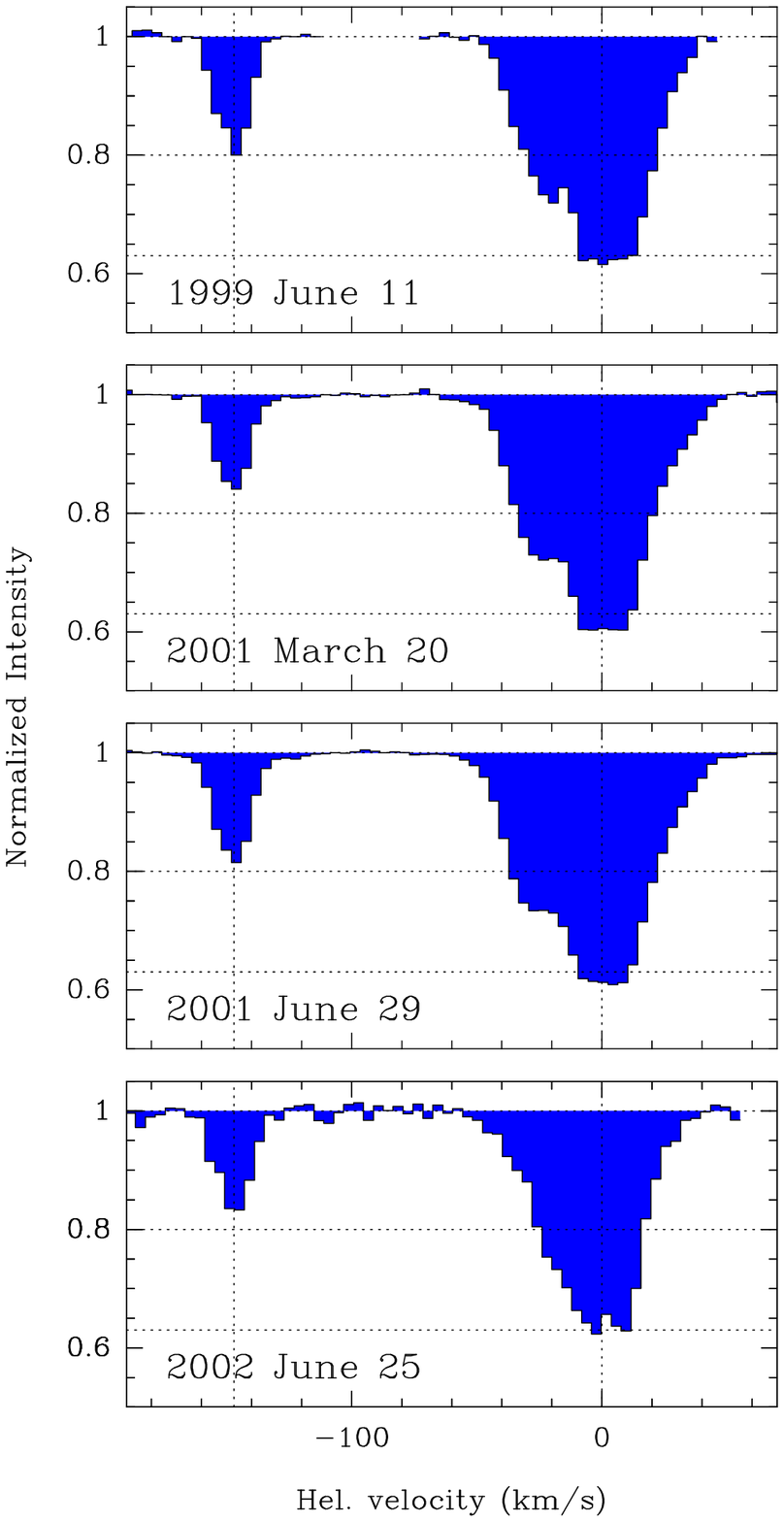}}
\caption{HCO$^+$ absorption lines observed at different epochs from 1999 to 2002.
For comparison, all spectra have been smoothed to the same spectral resolution of
1.25 MHz ($\simeq$ 4 kms$^{-1}$).}
\label{fig:hco-evol}
\end{figure}

In order to check for variations of $\Re$ during the period of our observations,
we have regularly observed the absorption spectrum of the HCO$^+$ ($2\leftarrow 1$)
line. \fig{hco-evol} shows the line profiles on 4 different epochs. We see no
variations in the lineshape or absorption depth that exceeds the 1-sigma rms
uncertainty (typically a few percent -- see \tab{magfactor}). The 0 kms$^{-1}$
component is known to be saturated in the HCO$^+$ line. Assuming all the emission
near 0 kms$^{-1}$ comes from the NE source, we derive relative intensities of
37\% and 63\% for the SW and NE continuum sources, and a magnification ratio		
$\Re$ = 1.66 (\tab{magfactor}). The lack of obvious variations of $\Re$ during our
observing period simply means that the flux variations over intervals of time of
25 days were either small, or constant.

During their 3-year long monitoring of the HCO$^+$ ($2\leftarrow 1$) line with
the 30-m telescope (1996 -- 98) Wiklind and Combes also observed magnification ratios
in the range 1.4 -- 1.7 with only a few excursions to values differing from $\Re$ = 1.6  
by more than 2$\sigma$ (1$\sigma$ was typically 0.1). Swift et al. (\cite{swi01}) reported
a significantly smaller ratio (1.2) for Dec. 1999, but this low ratio comes from the
separation of the two continuum sources in a half-resolved map. Our average ratio is
close to that measured with the VLA at 15 GHz and 23 GHz by Nair et al. (\cite{nai93}) in
1987 -- 1990 (1.5 -- 1.6). 

We therefore assumed in the following that the relative fluxes of the NE and SW
sources remained constant during our observations and simply co-added, when necessary,
the calibrated spectra observed on different sessions. A variation of $\Re$ (e.g.
by 20\%) for short periods of time is however not excluded: it would merely change
the absorption depths measured on those short periods (by $\simeq 13$\%). This would
have no significant effect on most isotopic ratios, that average data of different
observing periods (see \tab{obsdata}).

\begin{table}[h] \setcounter{Tableau}{3}
\caption{Measurements of the magnification factors.} \label{tab:magfactor} 
\begin{center} \begin{tabular}{ll}
\hline \hline
\multicolumn{1}{c}{Epoch} & \multicolumn{1}{c}{NE/SW ratio} \\
\hline
1995 Sep. 29 $^\dagger$ & 1.6 $\pm$ 0.2 \\
1995 Sep. 30 $^\dagger$ & 1.6 $\pm$ 0.2 \\
1995 Oct. 01 $^\dagger$ & 1.5 $\pm$ 0.1 \\
1995 Oct. 26 $^\dagger$ & 1.8 $\pm$ 0.2 \\
1999 Jun. 11 & 1.65 $\pm$ 0.02 \\
2001 Mar. 20 & 1.55 $\pm$ 0.04 \\
2001 Jun. 29 & 1.65 $\pm$ 0.02 \\
2002 Jun. 25 & 1.68 $\pm$ 0.05 \\
2003 Fev. 26 & 1.66 $\pm$ 0.07 \\
2003 Mar. 01 & 1.75 $\pm$ 0.07 \\
\hline
\end{tabular} \end{center}
\mbox{\,} %\vskip -.8cm
Notes: The magnification factor (NE/SW) is determined with the PdBI from
the saturation level of the HCO$^+$ ($2\leftarrow 1$)
line absorption at $V\simeq 0$ kms$^{-1}$. In the case of extended configuration,
baselines longer than 300 m were flagged out. \\
$^\dagger$ Re-reduction of previous PdBI observations (\cite{wc98}).
\end{table}

\subsection{Absorption profiles and opacities} \label{abs-profile}

\begin{figure}[h] \addtocounter{Figure}{1} \resizebox{\hsize}{!}
{\includegraphics{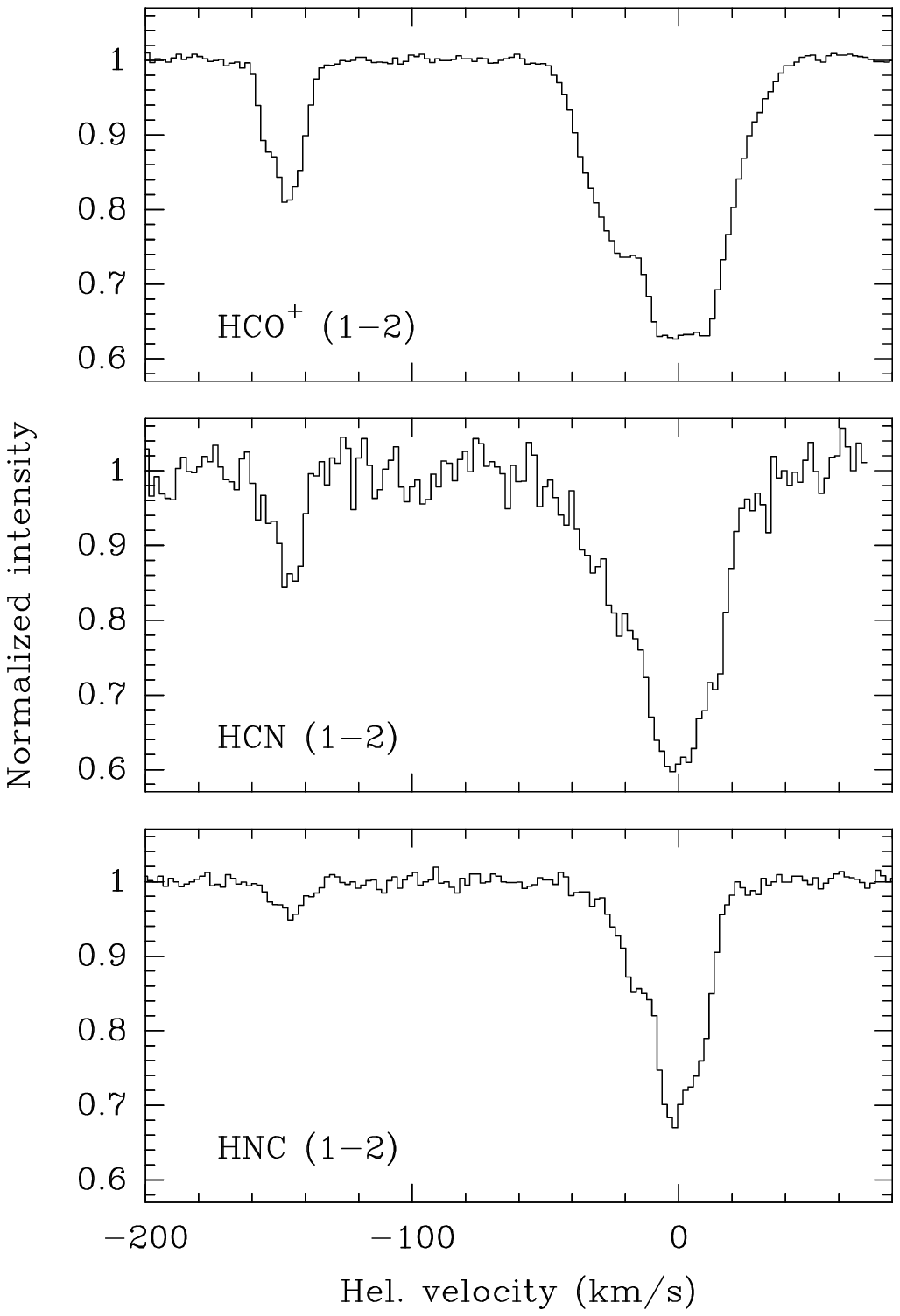}}
\caption{Spectra of the main isotopomers of HCO$^+$, HCN and HNC
(from top to bottom respectively). The velocity resolution is 2 kms$^{-1}$.}
\label{fig:spectra-strong}
\end{figure}

\fig{spectra-strong} presents the absorption profiles for the (J=$2 \leftarrow 1$)
lines of HCO$^+$, HCN and HNC. Both the V = 0 kms$^{-1}$ and V = $-$147 kms$^{-1}$ components
are detected in these 3 lines. The V = $-$147 kms$^{-1}$ component had been previously
observed by \cite{wc98}, but only for the HCO$^+$ line and with a lower signal to noise ratio.

The absorption profiles of the V = 0 kms$^{-1}$ and V = $-$147 kms$^{-1}$ components are not Gaussian.
The HNC ($2\leftarrow 1$) V = 0 kms$^{-1}$ component consists of at least 3 Gaussian
sub-components (\fig{show-tau}) and the V = $-$147 kms$^{-1}$ component of 2 sub-components,
which implies the presence of several clouds in front of the SW and NE sources. The 5
Gaussian sub-components have velocities of $\simeq$ $-$155, $-$150, $-$21, $-$3 and +7 kms$^{-1}$.
Their widths ($\simeq$ 10 -- 20 kms$^{-1}$) are typical of the width of Milky Way Giant Molecular Clouds.

A comparison of the HNC, HCN and HCO$^+$ line profiles suggests a similar set
of clouds with similar relative abundances, but with an increasing opacity
from HNC to HCO$^+$. A modelling of the line opacities, assuming that the SW
source contains 37\% of the total continuum flux, as derived in \S\ref{obs},
and that the source filling factor by the absorbing clouds is close to 1,
yields an opacity $\tau$ $\simeq$ 2 at the peak of the HNC line, and a HCN/HNC
opacity ratio of $\simeq$ 3, fairly constant across the entire spectrum, from
$-$160 to +20 kms$^{-1}$.

\begin{figure}[h] \addtocounter{Figure}{1} \resizebox{\hsize}{!}
{\includegraphics{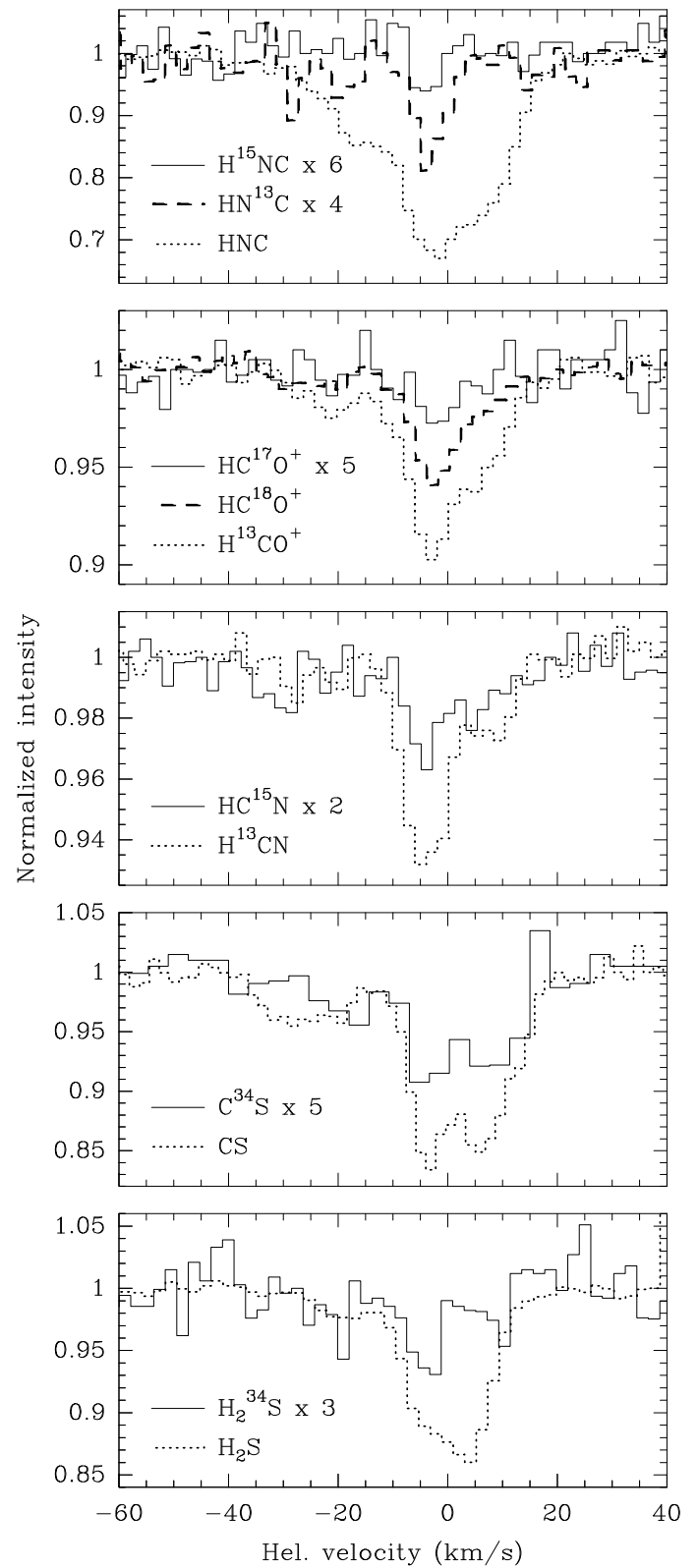}}
\caption{Spectra of the different species and isotopomers observed for this survey.
From top to bottom: HNC, HCO$^+$, HCN, CS and H$_2$S isotopomers. For a better
comparison, the weak isotopomers are scaled and renormalized, with scaling factor
indicated in each box. The velocity resolution is $\simeq$ 2 kms$^{-1}$ except for the
C$^{34}$S spectrum, for which it is $\simeq$ 4 kms$^{-1}$.}
\label{fig:spectra-weak}
\end{figure}

\begin{figure}[h] \addtocounter{Figure}{1} \resizebox{\hsize}{!}
{\includegraphics{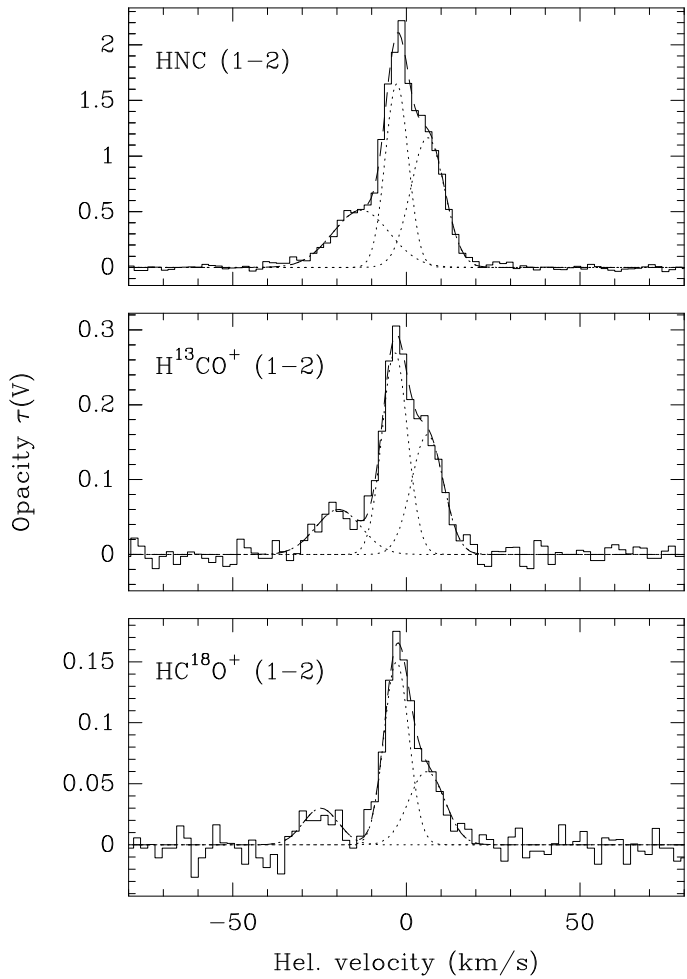}}
\caption{Line opacity profiles of the J=$2\leftarrow 1$ HNC (up), H$^{13}$CO$^+$ (middle)
and HC$^{18}$O$^+$ (down) for the SW component with the decomposition into
Gaussian components. Note the similarity of the profiles in spite of the large
difference in opacity.}
\label{fig:show-tau}
\end{figure}

\begin{figure}[h] \addtocounter{Figure}{1} \resizebox{\hsize}{!}
{\includegraphics{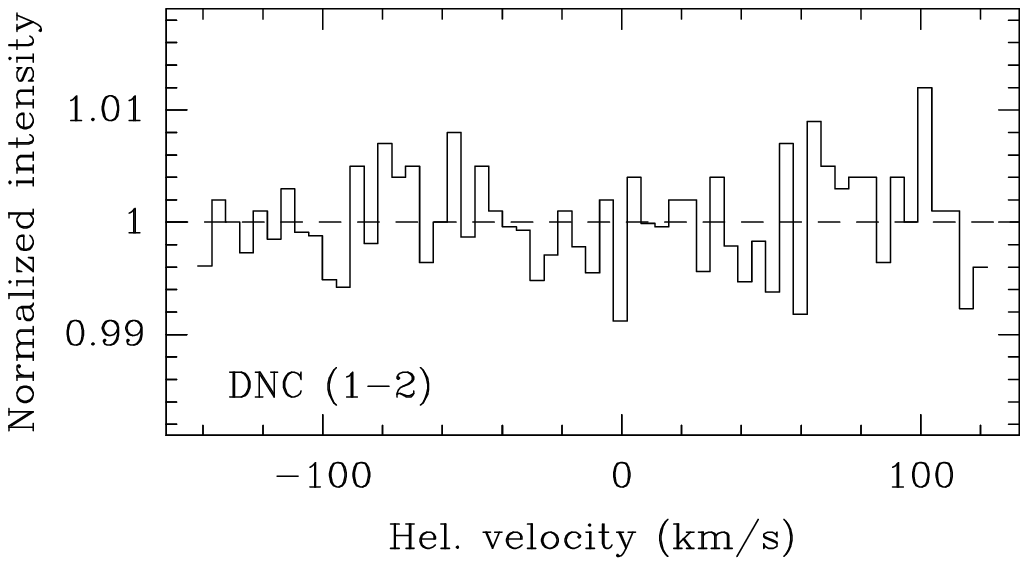}}
\caption{Spectrum observed at the frequency of the DNC ($2\leftarrow 1$)
line, smoothed to a resolution of 4.6 kms$^{-1}$. The RF was calibrated on the
quasar 3C454.3. No further baseline has been removed, except for a constant
offset. Note the quality of the baseline resulting from interferometric
observations. The narrow negative feature at 0 kms$^{-1}$ is not significant.}
\label{fig:dnc}
\end{figure}

\fig{spectra-weak} shows the J=$2\leftarrow 1$ line profiles of the rare $^{13}$C,
$^{15}$N, $^{17}$O and $^{18}$O isotopologues of HCO$^+$, HCN and HNC, as
well as those of CS and C$^{34}$S J=$4\leftarrow 3$, and H$_2$S and H$_2^{34}$S
(J$_{K_{-1},K_{1}}$)=1$_{10} \leftarrow 1_{01}$. For these species, only the SW
absorption is detected. The profiles of HNC and of the rare isotopomers
are not flat-topped, but show narrow peaks near 0 kms$^{-1}$. They are
not only sharper than those of the main HCO$^+$ and HCN isotopomers,
but also shallower, indicating that they are not saturated. The absorption
lines of H$^{13}$CO$^+$, HC$^{18}$O$^+$, H$^{13}$CN and HC$^{15}$N
show very similar profiles, while the spectra of CS and H$_2$S look
different: the +7 kms$^{-1}$ sub-component is much enhanced for those species. 
The similarity of the line opacity profiles of H$^{13}$CO$^+$ and HNC is
illustrated in \fig{show-tau}.

Assuming, again, that the SW source represents 37\% of the total flux
and that its filling factor is unity, the line opacity $\tau(v)$
is given by:
\begin{equation}
 \tau_{app}(v) = - \ln{ \left ( \frac{I(v)-I_{sat}}{|I_{sat}|} \right ) }
\end{equation}

\noindent where I(v) is the depth of the absorption from the continuum
level and I$_{sat}$ the saturation level: I$_{sat}^{SW}$ = $-$37\% and
I$_{sat}^{NE}$ = $-$63\% (we assume that the contribution from the Einstein
ring is negligible at mm wavelengths). We emphasize here that the knowledge
of the relative fluxes of the components allows us to derive the optical
depths without any assumption on the line excitation temperature other that
it is small with respect to the temperature of the quasar image.
The major source of uncertainty on $\tau$ is
coming from the filling factor, that may vary with velocity and species.
However, we expect that all the isotopologues have the same spatial
distribution, hence the same filling factor, as the abundant ones. 

The flat shape of the HCO$^+$ and HCN profiles around 0 kms$^{-1}$
suggests that the clouds cover entirely the SW source. This coverage,
however, may not be uniform, as the SW source size ($\geq$ 2.5 mas,
Carilli et al. \cite{car98}) is $\geq$ 10 pc at the distance of the galaxy and
presumably large with respect to individual clouds. The apparent
opacities in equation [1] are thus strictly speaking lower limits to
the true opacities. However, the similarity of the absorption
profiles on \fig{spectra-weak}, which yet have very different
opacities, suggests that the true opacities are not large, except for
HCO$^+$ and HCN near 0 kms$^{-1}$, and that the values derived from
equation [1] are close to the true opacities. This is illustrated in
\fig{show-tau} which shows the opacity profiles of the HNC,
H$^{13}$CO$^+$ and HC$^{18}$O$^+$ ($2\leftarrow 1$) lines, all of
which are indeed very similar.  We note that Carilli et al. (\cite{car98})
measure for the HCN ($1\leftarrow 0$) line an integrated apparent
opacity $\int \tau dV = 0.7$ kms$^{-1}$ for the NE (V = $-$147 kms$^{-1}$) component
that agrees fairly well with our value of 3.5 kms$^{-1}$ for the
($2\leftarrow 1$) line, demonstrating that the the component is indeed
optically thin.

We have fitted the opacity profiles (eg. \fig{show-tau}) derived with
equation [1] for all the observed species with 4 Gaussian components:
3 for the SW source ($-$50 $\leq$ V $\leq$ +40 kms$^{-1}$), and one for the
NE source ($-$160 $\leq$ V $\leq$ $-$140 kms$^{-1}$), which is too weak to be
resolved, except in the HCO$^+$ line. 
The fitted parameters (sub-component velocities, peak opacities and
velocity widths) and integrated opacities are listed in \tab{gaussfitSW}
and \tab{gaussfitNE} for the SW and NE components, respectively. The mean
central velocities of the 3 absorption components towards SW are $-$21,
$-$3 and 7 kms$^{-1}$, and their mean linewidths are 13, 8 and 10 kms$^{-1}$. The
linewidths of these different components are typical of Giant Molecular
Clouds in the Milky Way Molecular Ring.

Note that we could not derive reliable parameters for the rare isotopomer
HC$^{15}$N because of low S/N, and fixed the velocities of 2 Gaussian
profiles to $-$3 and 7 kms$^{-1}$. For the same reason, the NE absorption of the
CS spectrum was fitted with velocity and width fixed to the same values
as the HCN NE fit. 

\onltab{4}{
\begin{table*}[h] \setcounter{Tableau}{4}
\caption{Results of the decomposition of the SW opacity profiles into multiple Gaussian components.} \label{tab:gaussfitSW}
\begin{center} \begin{tabular}{lcccccccccccc}
\hline \hline
& 
\multicolumn{4}{c}{V $\simeq$ $-$3 kms$^{-1}$} &
\multicolumn{4}{c}{V $\simeq$ 7 kms$^{-1}$} &
\multicolumn{4}{c}{V $\simeq$ $-$23 kms$^{-1}$} \\
\multicolumn{1}{c}{SW} & 
V$_0$ & $\tau$ & $\Delta$V$_{1/2}$ & $\int \tau$dV & 
V$_0$ & $\tau$ & $\Delta$V$_{1/2}$ & $\int \tau$dV & 
V$_0$ & $\tau$ & $\Delta$V$_{1/2}$ & $\int \tau$dV \\
 & (kms$^{-1}$) & & (kms$^{-1}$) & (kms$^{-1}$) & & & (kms$^{-1}$) & (kms$^{-1}$) & & & (kms$^{-1}$) & (kms$^{-1}$) \\

\hline

H$^{13}$CO$^+$ (1-2)       & $-$3.3 & 0.27 & 8.6 & 2.5 (0.05)
                           & 6.1 & 0.16 & 10.8 & 1.8 (0.06) 
                           & $-$19.5 & 0.06 & 15.5 & 1.0 (0.07) \\

HC$^{18}$O$^+$ (1-2)       & $-$2.8 & 0.15 & 8.4 & 1.4 (0.06)
                           & 5.9 & 0.06 & 12.2 & 0.7 (0.07) 
                           & $-$24.6 & 0.03 & 11.8 & 0.3 (0.07) \\

HC$^{17}$O$^+$ (1-2)       & $-$1.8 & 0.01 & 10.0 & 0.15 (0.03)
                           & -- & -- & -- & --
                           & -- & -- & -- & -- \\

H$^{13}$CN (1-2)           & $-$3.8 & 0.21 & 9.2 & 2.1 (0.08)
                           & 8.4 & 0.08 & 7.6 & 0.6 (0.07)
                           & $-$25.6 & 0.03 & 11.0 & 0.3 (0.08) \\

HC$^{15}$N (1-2)           & $-$3.0 $^\dagger$ & 0.04 & 7.4  & 0.31 (0.03)
                           & 7.0  $^\dagger$ & 0.02 & 10.0 & 0.23 (0.04)
                           & -- & -- & -- & --  \\              

HNC (1-2)                  & $-$2.7 & 1.65 & 7.8 & 13.7 (0.07)
                           & 6.1 & 1.17 & 11.5 & 14.2 (0.08)
                           & $-$13.0 & 0.51 & 19.3 & 10.5 (0.1) \\

HN$^{13}$C (1-2)           & $-$3.3 & 0.12 & 6.8 & 0.9 (0.07)
                           & -- & -- & -- & -- 
                           & -- & -- & -- & -- \\

H$^{15}$NC (1-2)           & -- & (0.014)  & --  & $\leq$ 0.17
                           & -- & -- & -- & -- 
                           & -- & -- & -- & -- \\

DNC (1-2)                  & -- & (0.011) & -- & $\leq$ 0.19 
                           & -- & -- & -- & -- 
                           & -- & -- & -- & -- \\

CS (3-4)                   & $-$3.1 & 0.54 &  8.3 & 4.8 (0.08)
                           &   7.4 & 0.50 &  9.8 & 5.2 (0.09) 
                           & $-$25.5 & 0.13 & 16.3 & 2.3 (0.1) \\

C$^{34}$S (3-4)            & $-$3.5  & 0.06 & 7.4 & 0.44 (0.04) 
                           & 7.9   & 0.05 & 9.3 & 0.49 (0.05) 
                           & $-$18.1 & 0.02 & 10.1 & 0.25 (0.05) \\

H$_2$S (1$_{01}$-1$_{10}$) & $-$6.6 & 0.29 & 6.3 & 2.0 (0.05)
                           & 2.0 & 0.48 & 9.5 & 4.9 (0.06)
                           & $-$19.4 & 0.07 & 11.1 & 0.8 (0.07) \\

H$_2^{34}$S (1$_{01}$-1$_{10}$) & $-$4.3 & 0.06 & 7.2 & 0.49 (0.07) 
                                & -- & -- & -- & -- 
                                & -- & -- & -- & -- \\

H$_2^{33}$S (1$_{01}$-1$_{10}$) & -- & (0.016) & -- & $\leq$ 0.19
                                & -- & -- & -- & -- 
                                & -- & -- & -- & -- \\
\hline
\end{tabular} \end{center}
\mbox{\,} %\vskip -.8cm
Notes: Uncertainties on the integrated opacity are estimated as $\sigma _\tau \sqrt{\delta V
\Delta V_{1/2}}$ where $\sigma _\tau$ is the rms uncertainty on the opacity for the velocity
resolution $\delta$V ($\simeq$ 2 kms$^{-1}$ here).
Upper limits are calculated as 3$\sigma _\tau
\sqrt{\Delta V_{1/2} \delta V}$ where $\Delta$V$_{1/2}$ is taken as the velocity width of the
same component of the main isomer or isotopomer. 
The quoted uncertainties reflect the statistical noise and baseline residuals,
but not the uncertainties on the magnification ratio $\Re$, which is assumed to
remain constant. \\
$^\dagger$ Fixed parameter.
\end{table*}  
}

\begin{table}[h] \setcounter{Tableau}{5} 
\caption{Fit of the NE opacity profiles.} \label{tab:gaussfitNE}
\begin{center} \begin{tabular}{ccccc}
\hline \hline
Line            & V$_0$ & $\tau$ & $\Delta$V$_{1/2}$ & $\int \tau$dV \\
    & (kms$^{-1}$)& & (kms$^{-1}$) & (kms$^{-1}$) \\
\hline
HCO$^+$ (1-2) & $-$147.2 & 0.35 & 14.7 & 5.5 (0.05) \\
HCN (1-2)     & $-$146.4 & 0.28 & 11.6 & 3.5 (0.2) \\
HNC (1-2)     & $-$146.0 & 0.07 & 14.1 & 1.1 (0.05)\\
CS (3-4)      & $-$146.4 $^\dagger$ & 0.02 & 11.6 $^\dagger$ & 0.28 (0.08) \\
H$^{13}$CO$^+$ (1-2)  & -- & (0.007) & 14.7 $^\dagger$ & $\leq$ 0.12 \\
\hline
\end{tabular} \end{center}
\mbox{\,} %\vskip -.8cm
Notes: Uncertainties and upper limits on the integrated opacities are calculated as mentioned in
the caption of \tab{gaussfitSW}.\\
$^\dagger$ Fixed parameter.
\end{table}

\section{Isotopic ratios} \label{isotopicratios}

In order to derive the line opacity ratios and the column density ratios of
the different isotopologues, we assume that all the isotopologues from a same
molecular species have the same spatial distribution and are similarly excited,
hence that their line opacity profiles are similar.

The first hypothesis should be fulfilled, except for the species deeply
affected by isotopic fractionation. In cold Galactic clouds, the abundances
of molecules bearing $^{13}$C and, mostly, D atoms are known to be significantly
affected by fractionation reactions and selective photodissociation. This is not
the case, however, for the other species (see e.g. Langer et al. \cite{lan84}), and we
will assume that their relative abundances do reflect the elemental isotopic
abundance ratios. The hypothesis of similar spatial distribution is not critical
in the sense that it yields isotopic ratios averaged over the (small) continuum
source, as long as the line is not too optically thick.

The hypothesis of similar excitation is certainly correct for most line
components and species: all the species studied here have large dipole moments
and are hard to excite. Since the bulk of the absorption caused by a cloud
$>10$ pc with N(H$_2$) = few$\times 10^{22}$ cm$^{-2}$ (Menten et al. \cite{men99}) is
likely to arise in a low density envelope (n $\leq$ few$\times 10^2$ cm$^{-2}$),
the molecules are essentially heated by the 5.1 K cosmic background, so that
T$_{\rm rot}$ $\simeq$ 5.1 K. The lines of the main HCO$^+$ and HCN isotopologues,
the opacity of which is large between $-$10 and +10 kms$^{-1}$, were not considered
inside this velocity range for the derivation of the isotopic ratios.

The line opacity ratios derived for the different pairs of isotopologues are shown
in \tab{tau-ratio}. These ratios were calculated in the following way: five line
opacity profiles, assumed to be identical for all isotopologues of a given molecular
species, were derived by fitting the line opacity profiles of H$^{13}$CO$^+$,
H$^{13}$CN, HNC, CS and H$_2$S with Gaussian components, as explained in
\S\ref{abs-profile} (see \fig{show-tau} and \tab{gaussfitSW}). Those species
were selected because they are observed with a good signal-to-noise ratio and
have a low or moderate opacity. The fitted opacity profiles were then sampled at
the velocity resolution of the observations and squared to yield weights. The
rare-to-abundant isotope line opacity ratio was then calculated as the weighted
average $\bar{R}$ of the ratios measured for each velocity channel, in the velocity
ranges $-$10 $<$ V $<$ 10 kms$^{-1}$ (line center)
or $-$40 $<$ V $<$ $-$10 kms$^{-1}$ (line wing).
Finally, $\bar{R}$ was inverted to yield the abundant-to-rare line
opacity ratio, $\bar{R}^{-1}$.

Similarly, the uncertainties on the line opacity ratios were first estimated
for the rare-to-abundant ratios, $R$. In all cases, the uncertainties are
dominated by the noise on the rare isotopologue. Then, $R$ can be roughly decribed
as a normal variable of variance $\sigma^2$. The error bars on the inverse
(abundant-to-rare) ratio $\bar{R}^{-1}$ were then derived as equal to
$-(\bar{R}^{-1}-(\bar{R}+\sigma)^{-1})$ and +$((\bar{R}-\sigma)^{-1}-\bar{R}^{-1})$.
Note that the bars are asymmetric for low signal-to-noise ratios. A more accurate
evaluation of the errors from the likelihood function yields identical results.

Within our hypothesis of similar spatial distribution and similar excitation,
the values of \tab{tau-ratio}, multiplied by the inverse ratio of the square
of the line frequencies, reflect the molecular isotopic abundance ratios.

We note that the $^{12}$C/$^{13}$C and $^{14}$N/$^{15}$N ratios derived from
different species (e.g. HCN and HNC) and velocity components (e.g. wings {\em vs}
center of line) agree within the quoted uncertainties. We come back to this
point when we discuss the elemental isotopic ratios individually. 

\begin{table}[h] \setcounter{Tableau}{6}
\caption{Line opacity ratios derived from our observations with the IRAM interferometer.} \label{tab:tau-ratio}
\begin{center} \begin{tabular}{lcc}
\hline \hline
\multicolumn{1}{c}{Lines}           & Velocity & Line opacity    \\
			                    & interval &  ratios         \\
\hline
HCO$^+$ / H$^{13}$CO$^+$ ($2\leftarrow 1$)        & SW wings & 28 (3)  \\ 
HCN / H$^{13}$CN ($2\leftarrow 1$)                & SW wings & 40 ($-$5+7) \\
HNC / HN$^{13}$C ($2\leftarrow 1$)                & SW       & 27 (3)  \\
H$^{13}$CN / HC$^{15}$N ($2\leftarrow 1$)         & SW       & 4.2 (0.3) \\
HN$^{13}$C / H$^{15}$NC ($2\leftarrow 1$)         & SW       & 6 ($-$2+4) \\
HNC / H$^{15}$NC ($2\leftarrow 1$)                & SW       & 166 ($-$58+194) \\
HCO$^+$ / HC$^{18}$O$^+$ ($2\leftarrow 1$)        & SW wings & 53 ($-$10+16) \\
H$^{13}$CO$^+$ / HC$^{18}$O$^+$ ($2\leftarrow 1$) & SW       & 2.01 (0.07) \\
HC$^{18}$O$^+$ / HC$^{17}$O$^+$ ($2\leftarrow 1$) & SW       & 12 ($-$2+3) \\
CS / C$^{34}$S ($4\leftarrow 3$))                  &  SW      & 10.4 ($-$0.7+0.8) \\
H$_2$S / H$_2$$^{34}$S ($1_{10} \leftarrow 1_{01}$) & SW & 8 (1.5) \\
\hline
\end{tabular} \end{center}
\mbox{\,} %\vskip -.8cm
Notes: The SW component corresponds to velocities in the
range $-$10 to +10 kms$^{-1}$ and the SW wings to $-$40 to $-$10 kms$^{-1}$. The quoted
errors, 1$\sigma$, reflect the noise and baseline uncertainties, but
not the uncertainties on the magnification ratio $\Re$, which is assumed
to remain constant.
\end{table}

The elemental isotopic ratios derived by averaging
\footnote{The mean value of the $R_i,\sigma_i$ estimations was calculated as
$\bar{R}$ = ($\sum R_i/\sigma_i^2$)/($\sum 1/\sigma_i^2$) with a variance
$\sigma^2$ = 1/$(\sum 1/\sigma_i^2)$.} the values of \tab{tau-ratio} are given
in \tab{ratio}, where we list, for comparison, the isotopic ratios in a number
of Galactic and extragalactic sources. The former are the Solar System (SS),
whose isotopic ratios probably represent the state of the ISM 4.5 Gyr ago, the
local ISM, the Galactic Center (GC), which can be considered at a later stage
of evolution, and the carbon rich circumstellar envelope IRC+10216 that consists
of material entirely reprocessed in the core of a $\simeq$ 2 M$_\odot$ AGB star. The
latter are giant clouds from the Large Magellanic Could (LMC), that are
characterized by a low metallicity, and the nuclear regions of two nearby
starburst galaxies, NGC~253 and NGC~4945.

The local ISM values come from Lucas \& Liszt (\cite{luc98}) and were derived through
absorption measurements against extragalactic continuum sources. They probably
refer to the same type of clouds as those observed in our remote galaxy and are
the most accurate values of \tab{ratio}. The envelope IRC+10216 is probably
typical of the matter recycled to the ISM by low-mass stars. It is thought that two-third
of the matter presently ejected by the stars in the Galactic ISM comes from such
low mass stars. The sequence  SS-ISM-GC-IRC+10216 may thus be considered in the
first approximation as one of increasing processing of the gas by low-mass stars.

\onltab{7}{
\begin{table*}[h] \setcounter{Tableau}{7}
\caption{Comparison of the C, N, O and S isotopic ratios in the z = 0.89 galaxy (SW absorption) 
with different other environments.} \label{tab:ratio}
\begin{center} \begin{tabular}{lcccccc}
\hline \hline  
                     & $^{12}$C / $^{13}$C & $^{14}$N / $^{15}$N & $^{16}$O / $^{18}$O   & $^{18}$O / $^{17}$O & $^{32}$S / $^{34}$S \\ 
\hline
{\bf z = 0.89 galaxy}  & {\bf 27 $\pm$ 2}   & {\bf 130 $^{+20}_{-15}$} $^\dagger$     & {\bf 52 $\pm$ 4} $^\dagger$        & {\bf 12 $^{+3}_{-2}$}     & {\bf 10 $\pm$ 1}    \\ 
\hline
Solar System (a)     & 89                  & 270                       & 490                   & 5.5                 & 22                 \\  
Local ISM (b)        & 59 $\pm$ 2          & 237 $^{+27}_{-21}$        & 672 $\pm$ 110         & 3.65 $\pm$ 0.15       & 19 $\pm$ 8          \\ 
Galactic Center (c)  & 25 $\pm$ 5          & 900 $\pm$ 200             & 250 $\pm$ 30          & 3.5 $\pm$ 0.2       & 18 $\pm$ 5          \\ 
IRC+10216 (d)        & 45 $\pm$ 3          & $>$ 4400                  & 1260 $^{+315}_{-240}$ & 0.7 $\pm$ 0.2       & 21.8 $\pm$ 2.6      \\ 
\hline
LMC (e)              & 62 $\pm$ 5          & 114 $\pm$ 14              & $>$ 2000              & 1.8 $\pm$ 0.4       & 18 $\pm$ 6      \\    
NGC~253 (f)          & 40 $\pm$ 10         & --                        & 200 $\pm$ 50          & 6.5 $\pm$ 1         & 8 $\pm$ 2           \\ 
NGC~4945 (g)         & 50 $\pm$ 10         & 105 $\pm$ 25              & 195 $\pm$ 45          & 6.4 $\pm$ 0.3       & 13.5 $\pm$ 2.5      \\ 
\hline  
\end{tabular} \end{center}
\mbox{\,} %\vskip -.8cm
$^\dagger$ Derived from a double ratio assuming $^{12}$C/$^{13}$C = 27 $\pm$ 2. \\
References: 
a) Anders \& Grevesse \cite{and89};
b) Lucas \& Liszt \cite{luc98}, except for $^{18}$O/$^{17}$O taken from Penzias \cite{pen81};
c) Wilson \& Matteucci \cite{wil92}, Wilson \& Rood \cite{wil94} and references therein;
d) Kahane et al. \cite{kah88}, \cite{kah92} and Cernicharo et al. \cite{cer00};
e) Chin \cite{chi99a};
f) Henkel \& Mauersberger \cite{hen93}, Harrison et al. \cite{har99} and Mart\'in et al. \cite{mar05};
g) Wang et al. \cite{wan04}. 
\end{table*}
}

\subsection{Carbon}

The $^{12}$C/$^{13}$C elemental isotopic ratio is difficult to measure,
since the main $^{12}$C isotopologues tend to be saturated when its $^{13}$C
counterpart becomes detectable. This ratio, on the other hand, can be derived
in our survey from 3 different molecules: HCO$^+$, HCN and HNC. H$^{13}$CO$^+$,
H$^{13}$CN and HN$^{13}$C are all detected in the SW component (V = $-$3 kms$^{-1}$
component), but both HCO$^+$ and HCN are heavily saturated at the peak of the
line. The $^{12}$C/$^{13}$C ratio can thus be measured only in the line wings
(V $<$ $-$10 and V $>$ 10 kms$^{-1}$). The opacity of the HNC line ($\leq$ 2, see \fig{show-tau})
is low enough to allow an estimate of the $^{12}$C/$^{13}$C ratio even at the
center of the line. Within the uncertainties, we find the same ratio for the 3
molecular species: $\simeq 30$. The mean isotopic ratio, estimated from a weighted
average of the HCO$^+$, HCN and HNC line opacity ratios, is $27\pm 2$, where the
uncertainty only reflects the noise and the calibration errors, but not residual 
opacity effects linked e.g. to a change in the magnification ratio. Considering
these restrictions, our ratio is in good agreement with that derived by Menten et al. 
(\cite{men99}) from VLA HNC ($J=1\leftarrow 0$) line observations for the same velocity component:
$\simeq$ 35. The $J=1\leftarrow 0$ HNC line is less optically thick than the $J=2\leftarrow 1$ line, but
the VLA HN$^{13}$C spectrum has a lower velocity resolution and is noisier than
our spectra. 

The relative abundances of $^{12}$C and $^{13}$C can be affected by isotopic
fractionation (Lucas \& Liszt \cite{luc98}). Milam et al. (\cite{mil05}), however, argue that
the effects are limited and should not affect the isotopic ratios at the scales
involved here. The agreement between the ratios derived from three different
species and from two lines of the same species gives us confidence that the
$^{12}$C/$^{13}$C ratio in the SW source is indeed close to 27. Unfortunately,
none of the $^{13}$C species is detected in the NE component (V = $-$147 kms$^{-1}$).
The upper limit on H$^{13}$CO$^+$ yields a 3$\sigma$ lower limit to the 
HCO$^+$/H$^{13}$CO$^+$ ratio of 31, similar to the value of this ratio in the SW
component.

The interpretation of the $^{12}$C/$^{13}$C ratio in terms of nucleosynthesis
is also not straightforward. $^{12}$C is a primary product of helium burning in
intermediate and massive stars. $^{13}$C forms from $^{12}$C through the CNO cycles
in the H-burning shell. It may form in H-burning shell of the very massive stars
that produce $^{12}$C and/or could be produced later in a second generation of
intermediate or low mass stars (Prantzos et al. \cite{pra96}). In the later scheme, the
abundance of $^{13}$C relative to $^{12}$C is expected to increase with time and
the degree of processing of the gas, but the observed $^{12}$C/$^{13}$C ratio is
not significantly lower in the IRC+10216 envelope than in the local interstellar
medium (45 {\em vs} 59, see \tab{ratio}) which suggests that $^{13}$C behaves at least
partly as a primary element. Thus, the low value of the $^{12}$C/$^{13}$C ratio
in the z = 0.89 galaxy may not be very meaningful.

\subsection{Nitrogen}

A direct measurement of the $^{14}$N/$^{15}$N ratio is even more difficult
than for the $^{12}$C/$^{13}$C ratio, as it is larger. We derive this ratio
directly from HNC for the V = $-$3 kms$^{-1}$ component, or indirectly from the
double (H$^{13}$CN/HC$^{15}$N)x[$^{12}$C/$^{13}$C] ratio. The isotopic ratios
we arrive at are consistent and yield an average ratio $^{14}$N/$^{15}$N = 130
$\pm$ 20, a factor of two smaller than its Solar System and local ISM values.

The relative abundances of $^{14}$N and $^{15}$N are strongly affected
by nuclear processing. In low mass stars, where the temperature of the
H-burning zone is $\leq$ 10$^8$ K, the cold CNO cycles convert essentially
C, N and O into $^{14}$N and destroy most of $^{15}$N. In massive stars,
$^{15}$N is produced at equilibrium through the hot CNO cycle. In explosive
episodes that mark the end of those stars (novae and supernovae), $^{14}$N
is quickly converted into $^{15}$N and $^{18}$O. As a result, the
$^{14}$N/$^{15}$N ratio is a sensitive probe of the type of stars governing
the nucleosynthesis; it should increase with the degree of processing by
low mass stars. Indeed, $^{14}$N/$^{15}$N is found to be much larger in the
Galactic Center region than in the local ISM or in the Solar System and 
reaches an even larger value in the circumstellar shell IRC+10216
(\tab{ratio}). In contrast, the low ratio measured toward PKS~1830-211 fits
well with the scenario where low mass stars played almost no role in the
composition of the ISM in the z = 0.89 galaxy.

\subsection{Oxygen}

Because its value is very large, the $^{16}$O/$^{18}$O ratio is usually
measured through the double [$^{13}$C/$^{12}$C]x[$^{16}$O/$^{18}$O] ratio.
The value we derive in this way, $\simeq$ 60, is very low compared to the
Solar and interstellar values ($\simeq$ 600), even though allowance should
be made for a possible underestimation of $^{12}$C/$^{13}$C, due to
fractionation or residual opacity effects. That the actual ratio is low
is confirmed by our direct measurement of the H$^{12}$C$^{16}$O$^+$/H$^{12}$C$^{18}$O$^+$
ratio in the wings of the SW component (\tab{tau-ratio}), which also yields
$\simeq$ 60. This result is surprising, since both $^{16}$O and $^{18}$O
form in massive stars and are predicted to behave about similarly, except
at early ages, since $^{16}$O is a primary element and $^{18}$O a secondary
element. And indeed, the $^{16}$O/$^{18}$O ratio is found to decrease mildly
or not at all between the local ISM to the Galactic Center (see \tab{ratio}
and Polehampton et al. \cite{pol05}). At early ages, the $^{16}$O/$^{18}$O ratio is
supposed to be larger (Prantzos et al. \cite{pra96}). The $^{16}$O/$^{18}$O ratio in
the Solar System, 1.6 times smaller than in the ISM, is however a well known
puzzle in this context and is sometimes interpreted by an overabundance of
$^{16}$O in the Solar Nebula.

The $^{18}$O/$^{17}$O ratio, which is small and stems from lines that are
certainly optically thin, may be more accurately derived. According to the
current nucleosynthesis models, the $^{16}$O/$^{18}$O and $^{16}$O/$^{17}$O
isotopic ratios are sensitive indicators of the stellar mass, $^{18}$O being
essentially formed and released by short-lived massive stars while $^{17}$O 
comes mostly from long-lived low-mass stars (see e.g. Landr\'e et al. \cite{lan90},
Prantzos et al. \cite{pra96}). Our ratio $^{18}$O/$^{17}$O = 12 is very large,
compared to the canonical ISM value of 3 -- 4 (Penzias \cite{pen81}, Bensch et al. \cite{ben01},
Wouterloot et al. \cite{wou05}) and to the Solar System value of 5.5. It is especially
large with respect to the value observed in IRC+10216 and in other late type
giant envelopes which are all $<$ 1 (Kahane et al. \cite{kah92}). 

The high $^{18}$O/$^{17}$O and low $^{16}$O/$^{18}$O ratios, compared to the
ISM and, mostly, to IRC+10216, are probably linked to the relative youth of
the z = 0.89 galaxy. They may help us to better understand the puzzling $^{18}$O
nucleosynthesis. It is worth noting that Combes \& Wiklind (\cite{com95}) derived from
similar absorption observations a limit C$^{18}$O/C$^{17}$O $>$ 15 in another
high redshift galaxy (z = 0.68) on the line of sight to the quasar B0218+357.

The $^{18}$O/$^{17}$O ratio has been estimated in the starburst nuclei of M~82,
NGC~253 and NGC~4935 (Sage et al. \cite{sag91}, Henkel \& Mauersberger \cite{hen93}). The values,
which are much more uncertain than in the cases of the ISM and the z = 0.89 galaxy,
seem larger than the Local ISM value and may be explained by a initial mass
function biased towards massive stars (Henkel \& Mauersberger \cite{hen93}).

\begin{figure}[h] \addtocounter{Figure}{1} \resizebox{\hsize}{!}
{\includegraphics{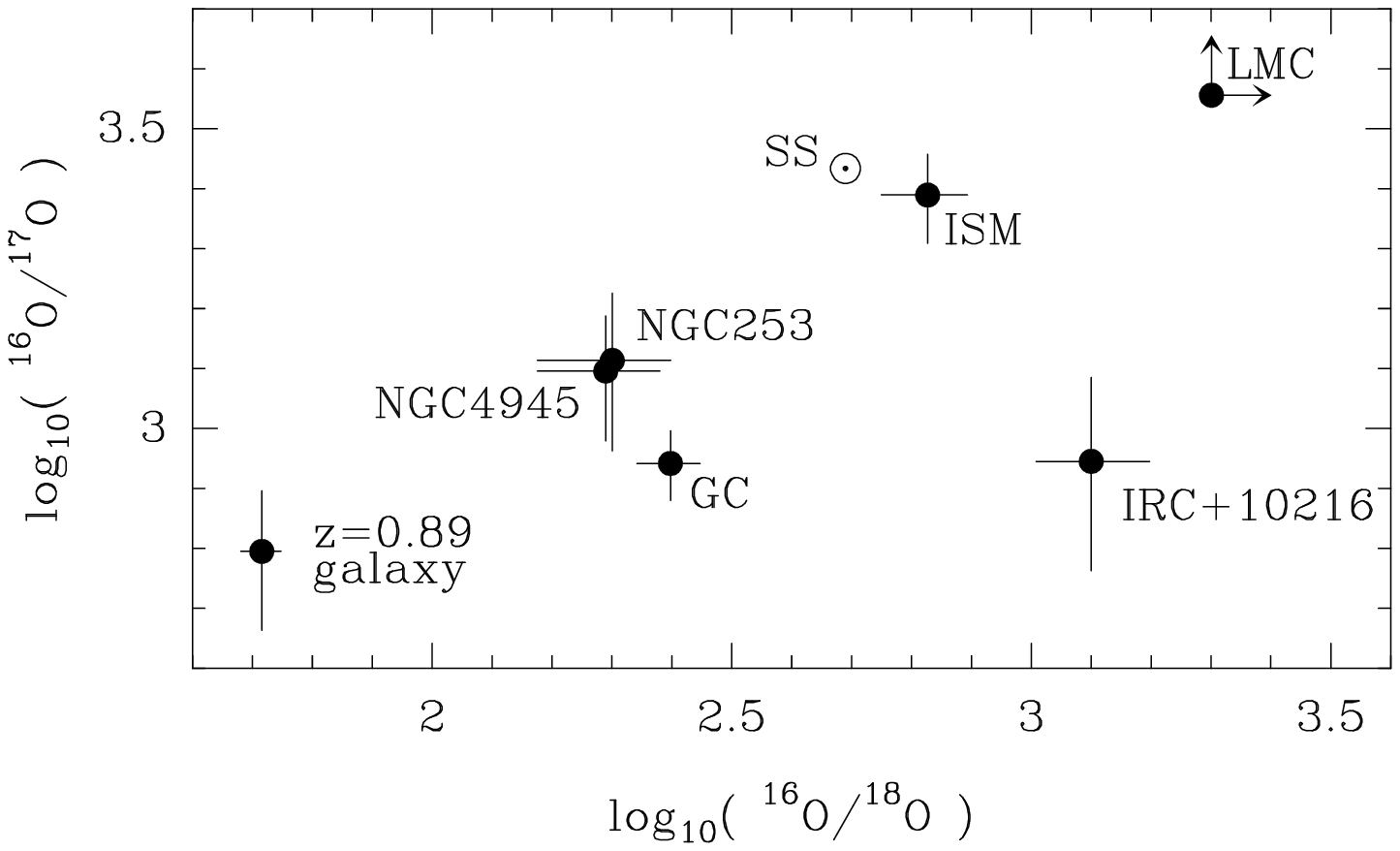}}
\caption{Oxygen isotopic ratios diagram $^{16}$O/$^{17}$O {\em vs} $^{16}$O/$^{18}$O
for different sources: the Solar System (SS), the Large Magellanic Cloud (LMC), the
local interstellar medium (ISM), the AGB star IRC+10216, the Galactic center (GC),
2 nuclei of nearby starbursts (NGC~253 and NGC~4945) and the z = 0.89 galaxy.
The values and references are listed in \tab{ratio}.}
\label{fig:o-diag}
\end{figure}

\fig{o-diag} compares in a single diagram the relative abundances of $^{16}$O,
$^{17}$O and $^{18}$O as observed for the main Galactic and extragalactic sources.
$^{17}$O appears particularly abundant in IRC+10216 and $^{16}$O unabundant in the
z = 0.89 galaxy. The LMC stands in this diagram at the opposite of the z = 0.89 galaxy, 
but this could be an artifact of low metallicity and strong UV radiation field, 
which may lead to selective photodissociation of the rare C$^{18}$O and C$^{17}$O
isotopes.

\subsection{Sulfur}

The $^{32}$S/$^{34}$S ratio we derive from CS and H$_2$S, 10 $\pm$ 1,
is at the same time accurate, since it is directly measured on optically
thin lines for two different species, and meaningful, since only few
molecular sources show such a clear departure of this ratio from its
Solar System value, 22. 

The $^{32}$S/$^{34}$S ratio is found to be the same in the processed 
stellar envelope IRC+10216, in the Solar System and in the local ISM.
The $^{34}$CS/$^{33}$CS ratio is the same in all Galactic molecular
sources, and the $^{13}$CS/C$^{34}$S ratio, which is used to estimate
the $^{32}$S/$^{34}$S ratio in Giant Molecular Clouds where $^{12}$CS
is optically thick, shows no variations across the Galaxy
(Frerking et al. \cite{fre80}, Chin et al. \cite{chi96}).

Because the $^{12}$C/$^{13}$C ratio derived from H$_2$CO and CO seems
to decrease between the Solar neighbourhood and the Galactic Center,
Chin et al. (\cite{chi96}) argue that the constancy of $^{13}$CS/C$^{34}$S
implies a similar decrease of $^{32}$S/$^{34}$S. The $^{12}$C/$^{13}$C
gradient, however, is shallow, except perhaps near the Galactic Center.
Moreover, CO and H$_2$CO yield for the same galactocentric radius discrepant
values (Wilson \& Rood \cite{wil94}). Thus, the observed $^{12}$CO/$^{13}$CO and
H$_2^{12}$CO/H$_2^{13}$CO ratios may well be affected by $^{13}$C fractionation
and/or line opacity effects and may not reflect the $^{12}$CS/$^{13}$CS ratio,
so that the presence of a $^{32}$S/$^{34}$S gradient in the Galaxy is not
established. As a matter of fact, it would be quite surprising to find a low 
$^{32}$S/$^{34}$S ratio in the Galactic Center, where nucleosynthesis,
we believe, is mainly driven by low mass stars. 

The sulfur isotopes are produced and destroyed in massive and very massive
stars. Their abundances stem from a critical balance between equilibrium
and explosive nucleosynthesis. In a 25 M$_\odot$ star, according to
Woosley \& Hoffman (\cite{woo86}),
$^{34}$S is 20 times more efficiently produced at equilibrium than $^{32}$S
and 100 times more than $^{33}$S. In an explosive oxygen burning episode,
the three isotopes are produced about evenly. Our $^{32}$S/$^{34}$S ratio
of 10, twice smaller that in the local ISM, could imply that the IMF in the
z = 0.89 galaxy is different from that of the Milky Way at the same age.
Unfortunately, our non-detection of H$_2^{33}$S brings little information
on this respect.   

Wang et al. (\cite{wan04}) and Mart\'in et al. (\cite{mar05}) report measurements of the
$^{32}$S/$^{34}$S ratio in the nuclear regions of two starburst galaxies,
NGC~4945 and NGC~253. The values they derive (13.5 $\pm$ 2.5 and 8 $\pm$ 2, 
respectively) are similar to that we find in the z = 0.89 galaxy. They are
much more uncertain, since they rely on the integrated intensity of the
broad and weak $^{13}$CS line and on the knowledge of the $^{12}$C/$^{13}$C
ratio. In contrast, the $^{32}$S/$^{34}$S in the LMC (18 $\pm$ 6, according to
Chin \cite{chi99a}) is close to that in the local ISM.

\subsection{Deuterium}

Deuterium is produced in the Big Bang nucleosynthesis with a primordial
[D/H] ratio of 2.5 10$^{-5}$ (Spergel et al. \cite{spe03}). It is completely
destroyed in the interior of stars, so that, ignoring infall of unprocessed
gas, the D/H ratio is an indicator of the degree of evolution. In the local
ISM, the average ratio D/H is $\sim$ 1.5 10$^{-5}$.

The deuterium abundance can be greatly enhanced in molecules in the 
cold and dense gas. Extensive modelling of D-chemistry (see e.g.
Roberts \& Millar \cite{rob00}) predicts D/H values as large as 0.01 -- 0.1 for
molecules embedded in very cold clouds, and such high ratios have been
observed in Galactic dark clouds (Gu\'elin et al. \cite{gue82}), pre-stellar cores
(Parise et al. \cite{par02}) and in the LMC (Chin et al. \cite{chi96}). 

We searched during 3 hours for the DNC  $J=2\leftarrow 1$ line with the PdB
interferometer; no absorption was detected down to a level of $\tau$ = 0.011
(3$\sigma$) with a velocity resolution of $\delta$V = 4.6 kms$^{-1}$. Assuming a
linewidth similar to the main isotopomer HNC, we obtain an upper limit from
the ratio DNC($2\leftarrow 1$)/HNC($2\leftarrow 1$): D/H $<$ 0.015 for the
$-$3 kms$^{-1}$ component. 

Our upper limit is less constraining than that derived by Shah et al. (\cite{sha99})
from the double isotopic ratio: [DCO$^+$]/[H$^{13}$CO$^+$].[$^{13}$C/$^{12}$C]
$<$ 0.0022. The latter, however, contrary to our result, depends on the
knowledge of the C isotopic ratio, a ratio that can be affected by
fractionation. Both limits remain far higher than the primordial ratio and
tell us little about stellar processing.

\section{Molecular column densities}

We have argued in \S\ref{isotopicratios} that the molecules studied here,
which all have a large dipole moment, are likely to be at equilibrium with
the cosmic background radiation, whose temperature at z = 0.89 is 5.1 K. The
observations support this conclusion: Menten et al. (\cite{men99}) measure a rotation
temperature T$_{\rm rot}$ = 4.5$_{-0.6}^{+1.5}$ K from the observations of two 
rotational transitions of HC$_3$N, and \cite{wc96} from
similar measurements find T$_{\rm rot}$ = 4 $\pm$ 2 for CS, H$^{13}$CO$^+$ and
N$_2$H$^+$ and T$_{\rm rot}$ $\leq$ 6 K for HNC. We therefore assume T$_{\rm ex}$ =
5.1 K in our calculations.

The column densities of \tab{ncol} were calculated from the equation:
\begin{equation}
\mathcal{N} = \frac{8 \pi \nu^3}{c^3 A_{ul} g_u} \frac{ Q(T_{\rm ex})
 \exp{(E_J/T_{\rm ex})}} {(1-\exp{(-h \nu / k T_{\rm ex})})} \int \tau dV 
\end{equation}

\noindent where we assume LTE, optically thin lines and a filling factor of unity. 
E$_J$ in this equation is the energy of the lower level of the transition, 
Q(T$_{\rm ex}$) = $\sum_{Ji} g_I(2Ji+1)\exp{(-E_{Ji}/kT_{\rm ex})}$ the partition
function (g$_I$ = 3 for an ortho level, = 1 in any other case), and A$_{ul}$ the
Einstein coefficient of the line
A$_{ul}$ = $\frac{64 \pi^4 \nu^3 \mu^2}{3 h c^3} \frac{S_{ul}}{g_u}$.
Here, $\mu$ is the electric dipole moment, S$_{ul}$ the line strength and
g$_u$ the degeneracy factor of the upper level.

As concerns HCO$^+$ and HCN, whose lines are optically thick, we calculated
their column densities by multiplying those of their $^{13}$C isotopologues
by $^{12}$C/$^{13}$C = 27. This assumes that the $^{12}$C/$^{13}$C ratio is
the same for all molecules and all velocity components. Our results are in
good agreement with the values of Menten et al. (\cite{men99}), which were derived
from rotational transitions with lower J and lower opacities.

\onltab{8}{
\begin{table*}[h] \setcounter{Tableau}{8} 
\caption{Total molecular column densities towards the SW and NE absorption.} \label{tab:ncol}
\begin{center} \begin{tabular}{ccccccccc}
\hline \hline
               & $\mu$   & S$_{ul}$ & g$_u$ & A$_{ul}$      & E$_{J,low}$ & Q(5.14 K) & $\mathcal{N}_{SW}$  & $\mathcal{N}_{NE} $ \\
               & (Debye) &          &       & (10$^{-4}$ s$^{-1}$)    & (K)  & & (10$^{12}$ cm$^{-2}$) & (10$^{12}$ cm$^{-2}$) \\
\hline

HCO$^+$        & 3.90    & 2        & 5     & 4.02  & 4.3   & 2.77    & 300 (23) $^\dagger$   & 11.4 (0.1)  \\
H$^{13}$CO$^+$ & 3.90    & 2        & 5     & 3.70  & 4.2   & 2.83    & 11.1 (0.2)    & $\leq$ 0.25   \\
HC$^{18}$O$^+$ & 3.90    & 2        & 5     & 3.50  & 4.1   & 2.88    & 5.1 (0.3)         & --   \\
HC$^{17}$O$^+$ &         &          &       &       &       &         & 0.4 (0.1) $^\dagger$  & -- \\

HCN            & 2.99    & 2        & 5     & 2.32  & 4.3   & 2.78    & 289 (25) $^\dagger$ & 12.3 (0.7) \\
H$^{13}$CN     & 2.99    & 2        & 5     & 2.14  & 4.1   & 2.84    & 10.7 (0.5)           & $\leq$ 0.5  \\
HC$^{15}$N     & 2.99    & 2        & 5     & 2.13  & 4.1   & 2.85    & 2.1 (0.1)            & --  \\

HNC            & 3.05    & 2        & 5     & 2.58  & 4.4   & 2.73    & 129.0 (0.5)         & 3.7 (0.2) \\
HN$^{13}$C     &     &        &     &  &   &    & 4.8 (0.4)$^\dagger$    & --   \\

H$^{15}$NC     &     &         &     &   &   &     & 1.0 (0.2) $^\dagger$ &  --  \\
DNC            & 3.05    & 2        & 5     & 1.54  & 3.7   & 3.17    & $\leq$ 0.7             & --  \\

CS             & 1.96    & 4        & 9     & 1.50  & 14.1  & 4.72    & 560 (7)            & 13 (4)  \\
C$^{34}$S      & 1.96    & 4        & 9     & 1.42  & 13.9  & 4.79    & 53 (4)             & --    \\

H$_2$S         & 0.97    & 4.5      & 9     & 0.26  & 19.8  & 1.28   & 1104 (15)   & $\leq$ 17 \\
H$_2^{34}$S    &     &       &     &   &   &     & 110 (11) $^\dagger$ & --   \\
\hline
\end{tabular} \end{center}
\mbox{\,} %\vskip -.8cm
Notes: The column densities marked with $^\dagger$ were estimated using the value of a nearby isotopomer,
corrected from the corresponding isotopic ratio. Upper limits on column densities are
calculated from the 3$\sigma$ upper limits on the integrated opacities in \tab{gaussfitSW}
and \tab{gaussfitNE}.
\end{table*}  
}

A crude analysis of the chemistry in the z = 0.89 galaxy has been made 
by \cite{wc96} and Menten et al. (\cite{men99}), who concluded that
the molecular abundances in the SW velocity component (V $\simeq$ 0 kms$^{-1}$)
are similar to those of Galactic dark clouds such as TMC~1. The HNC/HCN
ratio we find (0.4) is indeed similar to that in those cold clouds
(Hirota et al. \cite{hir98}) and two orders of magnitude larger than in the hot
and dense clouds, such as Orion MC-1. A high HNC/HCN ratio probably means
that these two species are formed by gas phase ion-molecule reactions via
HCNH$^+$. Note that a high HNC/HCN ratio has also been reported in the
center of nearby starburst galaxies (e.g. Mart\'in et al. \cite{mar06}).

The HCO$^+$/N$_2$H$^+$ ratio, which reflects in the medium dense gas
the CO/N$_2$ ratio, is 6 (adopting the N$_2$H$^+$ integrated ($2\leftarrow 1$)
line opacity measured by \cite{wc96}). Its value is close to the $''$cosmic$''$
O/N elemental ratio, 7.

Until now, H$_2$S extragalactic detection was only reported in the LMC
(Heikkil\"a  et al. \cite{hei99}) and NGC~253 (Mart\'in et al. \cite{mar05}). The abundance of
H$_2$S in the SW component is large (few $\times 10^{-8}$ with respect to
H$_2$, adopting $\mathcal{N}$(H$_2$) = 3 10$^{22}$ cm$^{-2}$, \cite{wc96}),
compared to the standard values of 1 -- 3 10$^{-9}$ measured in the quiescent
regions of OMC-1 and in the dark clouds (Minh et al. \cite{min89}). It is also a
factor of $\simeq$ 10 larger than in the local diffuse clouds, as measured
by Lucas \& Liszt (\cite{luc02}). This overabundance could reflect a large abundance
of S in the gas, although the CS abundance seems normal. It could also mean
that the absorbing gas at z = 0.89 is not that quiescent and that H$_2$S
molecules could be released from dust grains either by intense UV radiation
from star forming regions or by shocks generated by young stellar objects,
as is observed in the Orion (KL) outflow (Minh et al. \cite{min90}). The very wide
absorption profile observed in a few pc-large SW absorbing cloud may
support this last scenario.  

Although we have measured only a few column densities for the NE velocity
component (V = $-$147 kms$^{-1}$), those also yield relative abundances similar
to those of the cold dark clouds. In particular, the HNC/HCN ratio (0.3)
is large.

\section{Conclusion}

The presence of an intervening nearly face-on spiral galaxy at z = 0.89
in the line of sight of the radio loud quasar PKS~1830-211 offers an
unique chance to study the chemical and isotopic compositions of the
interstellar medium in the spiral arm of a galaxy 2 -- 3 times younger than the
Milky Way. The use of the PdB interferometer has enabled us to combine
very flat baselines and high sensitivity, making it possible to measure
reliably the C, N, O and S isotopic abundance ratios in HCO$^+$, HCN,
HNC, CS, and H$_2$S. The ratios derived from different species or velocity
components are fully consistent, which supports the view that they are not
affected by residual opacity effects and/or fractionation effects and that
they reflect the elemental isotopic ratios.

A comparison of the isotopic ratios in the z = 0.89 galaxy with those
observed in the local ISM, the Galactic Center, the circumstellar
envelope IRC+10216 and the Solar System shows major differences with 
these sources. Mainly, the $^{17}$O/$^{18}$O, and $^{14}$N/$^{15}$N
ratios, which are sensitive probes of the degree of processing of the
gas by low mass stars and which are expected to increase with time are
found to be smaller than in the ISM or the Solar System and much
smaller than in the Galactic Center and in the fully processed
low-mass star envelope IRC+10216. Most remarkably, the observed
$^{16}$O/$^{18}$O ratio is a factor of 10 lower than in the ISM and
the Solar System and the $^{32}$S/$^{34}$S ratio is twice lower
than in the Galactic sources. These latter ratios result mainly from 
nucleosynthesis in high mass stars, it is likely that the stellar
Initial Mass Function in the z = 0.89 galaxy is different from that of
the Milky Way at the same age. The low $^{17}$O/$^{18}$O, and
$^{14}$N/$^{15}$N ratios are consistent with the young age of the
galaxy ($<$ 6 Gyr) which is smaller than the lifetime of stars with
masses $<$ 1.5 M$_\odot$ and severly limits the importance of low mass
stars in the regeneration of the ISM.

The isotopic ratios observed towards PKS~1830-211, although more extreme,
are not unsimilar to those derived in the starburst nuclei of nearby
galaxies, such as NGC~253, M82 and NGC~4945 (see e.g. Mart\'in et al. \cite{mar06}).
The later ratios, which result from the observation of weak and broad
emission lines, are however far more uncertain.

The extension of this study to other remote galaxies is hampered by the small
number of millimeter radio sources whose line-of-sight intercepts the
plane of a remote galaxy. PKS~1830-211 is by far the brightest such source.
The second best, at present, is the gravitational lens B0218+357 and
several molecular lines have been observed in absorption at z = 0.68
toward this object (Wiklind \& Combes \cite{wik95}, Menten \& Reid \cite{men96},
Combes \& Wiklind \cite{com97}). Even in this object, the detection of less
abundant species is difficult (Henkel et al. \cite{hen05}). Fortunately, the
flux of this source has been growing lately at millimeter wavelengths,
giving hope that the rare carbon, oxygen and sulfur isotopes can be
studied in the near future.

\begin{acknowledgements}
We would like to thank all the members of the IRAM staff from the Plateau de Bure
who made these observations possible. S.M. thanks Dr. Y.C. Minh for fruitful and stimulating
discussions. We wish to thank the referee, whose comments helped to improve the paper.
Based on observations carried out with the IRAM Plateau de Bure Interferometer.
IRAM is supported by INSU / CNRS (France), MPG (Germany) and IGN (Spain).
\end{acknowledgements}

\end{document}